\documentclass[11pt,a4paper]{article}
\usepackage{amsfonts}
\usepackage{amssymb}
\usepackage{amsmath}
\usepackage{graphicx}
\usepackage{booktabs}
\usepackage{ulem}
\usepackage[small,bf]{caption}
\usepackage{cite}
\usepackage{color}
\usepackage{a4wide}

\allowdisplaybreaks[1]

\newcommand{\be}{\begin{equation}}
\newcommand{\ee}{\end{equation}}
\newcommand{\beq}{\begin{equation}}
\newcommand{\eeq}{\end{equation}}
\newcommand{\bea}{\begin{eqnarray}}
\newcommand{\eea}{\end{eqnarray}}

\newcommand{\martin}[1]{{\color{blue} #1}} 

\begin{document}

\begin{titlepage}

\vspace*{-15mm}
\begin{flushright}
MPP-2013-282\\
SISSA 49/2013/FISI\\
TTP13-035
\end{flushright}
\vspace*{0.7cm}

\begin{center}
{
\bf\LARGE
GUT predictions for quark-lepton Yukawa coupling  \\[1mm]
ratios with messenger masses from non-singlets
}
\\[8mm]
Stefan~Antusch$^{\star}$
\footnote{E-mail: \texttt{stefan.antusch@unibas.ch}},
Stephen~F.~King$^{\dagger}$
\footnote{E-mail: \texttt{king@soton.ac.uk}},
Martin~Spinrath$^{\ddag}$
\footnote{E-mail: \texttt{martin.spinrath@kit.edu}},
\\[1mm]
\end{center}
\vspace*{0.50cm}
\centerline{$^{\star}$ \it
 Department of Physics, University of Basel,}
\centerline{\it
Klingelbergstr.~82, CH-4056 Basel, Switzerland}
\vspace*{0.2cm}
\centerline{$^{\star}$ \it
Max-Planck-Institut f\"ur Physik (Werner-Heisenberg-Institut),}
\centerline{\it
F\"ohringer Ring 6, D-80805 M\"unchen, Germany}
\vspace*{0.2cm}
\centerline{$^{\dagger}$ \it
School of Physics and Astronomy, University of Southampton,}
\centerline{\it
SO17 1BJ Southampton, United Kingdom }
\vspace*{0.2cm}
\centerline{$^{\ddag}$ \it
SISSA/ISAS and INFN,}
\centerline{\it
Via Bonomea 265, I-34136 Trieste, Italy }
\vspace*{0.2cm}
\centerline{$^{\ddag}$ \it
Institut f\"ur Theoretische Teilchenphysik, Karlsruhe Institute of Technology}
\centerline{\it
Engesserstra\ss{}e 7, D-76131 Karlsruhe, Germany }
\vspace*{1.20cm}
\begin{abstract}
\noindent
We propose new predictions from Grand Unified Theories (GUTs)
(applicable to both supersymmetric (SUSY) and non-SUSY models)
for the ratios of quark and lepton Yukawa couplings.
These new predictions arise from splitting the masses of the messenger fields for
the GUT scale Yukawa operators by Clebsch-Gordan factors from GUT symmetry breaking.
This has the effect that these factors enter inversely in the predicted quark-lepton
Yukawa coupling ratios, leading to new possible GUT predictions.  We systematically construct
the new predictions that can be realised in this way in SU(5) GUTs and Pati-Salam
unified theories and discuss model building applications. 
\end{abstract}

\end{titlepage}

\setcounter{footnote}{0}

\section{Introduction}

In unified theories of fermion masses and mixings, such as in SU(5) Grand Unified Theories (GUTs) \cite{Georgi:1974sy}
or Pati-Salam (PS) models \cite{Pati:1974yy}, the fermions of the Standard Model (SM) (plus in the latter case
right-handed neutrinos) are unified in joint representations of the enlarged unified gauge
symmetry group. When this symmetry gets broken to the SM, this can result in
predictions for the ratios of the entries of the quark and lepton Yukawa matrices at the unification scale
$M_\mathrm{GUT}$. Typical examples for such predictions are bottom-tau Yukawa coupling unification
\cite{Georgi:1974sy}, i.e.\ $y_b = y_\tau$, at $M_\mathrm{GUT}$, or the Georgi-Jarlskog
relation $y_\mu/y_s = 3$ \cite{GJ}. Both ratios can emerge from renormalisable GUT operators
for Yukawa couplings in the context of SU(5) GUTs or Pati-Salam models
(seen as a step towards SO(10) GUTs). The phenomenology of these relations was studied
quite extensively, for some recent references, see, for instance, \cite{Ross:2007az,
Altmannshofer:2008vr, Baer:2009ff, Gogoladze:2009ug, Anandakrishnan:2012tj}.

However, in models for fermion masses and mixings which aim at explaining the hierarchies
between the quark and charged lepton Yukawa couplings of the different generations, the
Yukawa couplings smaller than ${\cal O}(1)$ are preferably generated by effective operators,
realised by the exchange of heavy messenger fields in GUT representations. This applies to
the Yukawa couplings of the first and second generations, but also to the bottom and tau Yukawa
couplings in supersymmetric (SUSY) models with small or moderate $\tan \beta$ 
or in non-SUSY models. In the case of SUSY models, we assume that the low energy theory
corresponds to the minimal supersymmetric standard model (MSSM).

The new possibilities which arise from such operators have been discussed in the context of
SU(5) GUTs in \cite{Antusch:2009gu} and in Pati-Salam models in \cite{Allanach:1996hz}, 
\cite{Allanach:1997gu} and
\cite{Antusch:2009gu} with dimension 6 and dimension 5 operators, respectively.
For similar studies in SO(10) see, for instance, \cite{Anderson:1993fe} or more recently
\cite{Bazzocchi:2008sp}. It has been shown in \cite{Antusch:2009gu} that in
many SUSY scenarios the new SU(5) relations like $y_\tau/y_b = -\tfrac{3}{2}$ or
$y_\mu/y_s = \tfrac{9}{2}$ or $6$ are often favoured compared to bottom-tau
Yukawa unification or the Georgi-Jarlskog relation. The implications of these new relations
were studied, for instance, in \cite{Monaco:2011wv, Antusch:2011sq, Antusch:2011xz}. 
In any case, towards making progress in building SUSY or non-SUSY GUT models of flavour, it is
important to be aware of the full set of possibilities
and to study their phenomenological consequences. 

In this paper, we discuss a new way of obtaining predictions for the GUT scale ratios of quark
and lepton Yukawa couplings. These new possibities arise from splitting the masses of the
messenger fields for the GUT scale Yukawa operators by Clebsch-Gordan (CG) factors from GUT
symmetry breaking. This has the effect that these factors enter inversely in the predicted
quark-lepton mass relations, leading to new possible predictions.  We systematically
list the new predictions that can be realised in this way in SU(5) GUTs and Pati-Salam unified
theories and discuss model building applications.

\section{Predictions from SU(5) Unification}
\label{Sec:SU5}

\begin{figure}
\centering
\includegraphics[scale=1]{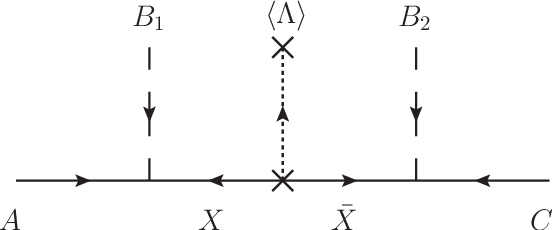}
\caption{
Diagrammatic representation of the operators giving
Yukawa coupling ratios at the GUT scale. $A$ and $C$
are matter fields while $B_1$ and $B_2$ are Higgs fields.
$\langle \Lambda \rangle$ represents the messenger mass
term. If $\Lambda$ is a total singlet one could directly
write down a mass like in \cite{Antusch:2009gu}. Otherwise
$\Lambda$ should transform as an adjoint of the GUT symmetry
splitting the masses of the components of the messenger fields.
\label{Fig:Dim5}
}
\end{figure}

In \cite{Antusch:2009gu} some of the authors considered SU(5) GUT mass ratios from
dimension five operators. The approach can be most easily explained looking at
Fig.~\ref{Fig:Dim5}. In the previous publication the vacuum expectation value (vev)
of $\Lambda$ was not considered or in other words the mass of the messenger fields pair
$X$ and $\bar X$ was assumed to transform trivially as a gauge singlet. The external fields $A$,
$B_1$, $B_2$ and $C$  (note that we use here a slightly different notation) were
assigned to various GUT representations. In particular, one was the five-dimensional
matter representation of SU(5)
\begin{equation}
F = \overline{\mathbf{5}} =  \begin{pmatrix}
         d_R^{c} & d_B^{c} & d_G^{c} & e &-\nu
          \end{pmatrix}_L \;, 
\end{equation}
one was the ten-dimensional matter representation of SU(5)
\begin{equation}
T = \mathbf{10}
        = \frac{1}{\sqrt{2}}
           \begin{pmatrix}
           0 & -u_G^{c} & u_B^{c} & -u_{R} & -d_{R} \\
           u_G^{c} & 0 & -u_R^{c} & -u_{B} & -d_{B} \\
           -u_B^{c} & u_R^{c} & 0 & -u_{G} & -d_{G} \\
           u_{R} & u_{B} & u_{G} & 0 & -e^c \\
           d_{R} & d_{B} & d_{G} & e^c & 0
           \end{pmatrix}_L \:
\end{equation}
and one was the SU(5) representation containing the Higgs doublet.
This can sit either in a five- or in a 45-dimensional representation
\begin{align}
 \left( \bar{h}_5 \right)^a = \overline{\mathbf{5}}^{\, a}  & \;, \quad  \langle \left( \bar{h}_5 \right)^5 \rangle = v_5 \;,\\
\left(\bar{h}_{45}\right)^{ab}_c = -\left(\bar{h}_{45}\right)^{ba}_{c} = \overline{\mathbf{45}}^{\, ab}_{\, c} &\;, 
\quad\langle \left(\bar{h}_{45}\right)^{i5}_j \rangle = v_{45} \left(\delta^i_j - 4 \delta^{i4} \delta_{j4}\right)\;,
\end{align}
where $a,b=1,\ldots,5$, $\alpha=1,2,3$, $\beta=4,5$ and $i,j=1,\ldots,4$. 
The fourth external field of the diagram in Fig.~\ref{Fig:Dim5} was
assigned to a GUT symmetry breaking Higgs field, like the adjoint 
\begin{equation}
\left(H_{24}\right)^a_b = \mathbf{24}^{\, a}_{\, b} \;, \quad 
\langle \left(H_{24}\right)^a_b \rangle = V_{24} (2 \delta^a_\alpha \delta_b^\alpha - 3 \delta^a_\beta \delta_b^\beta)\;, \label{Eq:H24vev} \\
\end{equation}
where again $a,b=1,\ldots,5$, $\alpha=1,2,3$ and $\beta=4,5$.

If we identify $A$ and $C$ with the matter representations $F$ and $T$, then
there are new possible Yukawa coupling relations beyond
the renormalizable ones. With the vev of the additional
GUT Higgs field $H_{24}$ pointing in the hypercharge direction we found, for instance, 
the new GUT predictions $(Y_e)_{ij}/(Y_d)_{ij} = -3/2$, $9/2$ and $6$, depending on
which representation the $H_{24}$ field couples to and on 
which representation contains the SM Higgs doublet.
In the SUSY case the fields $A$ 
and $C$ would be fermionic components and the fields $B_1$, $B_2$ and $\Lambda$ 
scalar components of the respective superfields. 
As already noted, our results also apply to the non-SUSY case.

In this work we want to extend the above approach to include the
case that the mass of the messenger pair $X$ and $\bar X$ is not  
generated by a gauge singlet $\Lambda_1$ but by a GUT non-singlet,
e.g.\ by a field in the adjoint representation of 
SU(5), $\Lambda_{24}$. In this particular case the masses
of the components of the messenger fields are proportional to their
respective hypercharges. More generally, when the heavy messenger fields (which in general have
masses split by CG factors according to our approach) are integrated out, these CG factors will enter
inversely in the considered down-type quark and charged lepton Yukawa
matrix elements, which can lead to new GUT predictions for the ratios
$(Y_e)_{ij}/(Y_d)_{ij}$ beyond the ones considered in \cite{Antusch:2009gu}.    
Note that in the diagram of Fig.~\ref{Fig:Dim5} the field $\Lambda$,
which acquires a vev, should not be viewed as an external field of an
effective operator. When $\Lambda$ acquires its vev, it generates the mass
term for the messenger pair $X$ and $\bar X$ and only then the messengers
can be integrated out to obtain an effective operator. Consequently, the
vev $\langle \Lambda \rangle$ will appear inversely in the effective operator.

\begin{table}
\begin{center}
\begin{tabular}{cc}\toprule
Operator Dimension & $(Y_e)_{ij}/(Y_d)_{ij}$ 	\\ 
\midrule 4	&	1	\\
	&	-3	\\
\midrule 5	&	{\color{red} 1/6}\\
	&	-1/2			\\
	&	{\color{red} -2/3}	\\
	&	1			\\
	&	$\pm$3/2		\\
	&	-3			\\
	&	9/2			\\
	&	6			\\
	&	-18			\\
\bottomrule
\end{tabular}
\end{center}
\caption{
Summary of possible SU(5) predictions for the GUT scale
Yukawa coupling ratios $(Y_e)_{ij}/(Y_d)_{ij} $. The new relations compared
to \cite{Antusch:2009gu} are shown in red. For more details
about which operator gives which ratio, see Tab.~\ref{Tab:SU5Messenger1}. 
\label{Tab:SU5Relations}}
\end{table}

\begin{table}
\begin{center}
\begin{tabular}{ccccc} \toprule
($A$, $B_1$) & ($C$, $B_2$) & $X$ & $\Lambda$ & $(Y_e)_{ij}/(Y_d)_{ij}$ \\ \midrule
($F$, $\bar{h}_5$) & ($T$, $H_{1}$) & $\mathbf{10}$ & $\Lambda_1$ & 1 \\
($F$, $\bar{h}_{45}$) & ($T$, $H_{1}$) & $\mathbf{10}$ & $\Lambda_1$ & -3 \\
($F$, $\bar{h}_5$) & ($T$, $H_{24}$) & $\mathbf{10}$ & $\Lambda_1$ & 6 \\
($F$, $\bar{h}_5$) & ($T$, $H_{24}$) & $\mathbf{15}$ & $\Lambda_1$ & 0 \\
($F$, $\bar{h}_5$) & ($T$, $H_{75}$) & $\mathbf{10}$ & $\Lambda_1$ & -3 \\
($F$, $\bar{h}_{45}$) & ($T$, $H_{24}$) & $\mathbf{10}$ & $\Lambda_1$ & -18 \\
($F$, $\bar{h}_{45}$) & ($T$, $H_{24}$) & $\mathbf{40}$ & $\Lambda_1$ & 0 \\
($F$, $\bar{h}_{45}$) & ($T$, $H_{75}$) & $\mathbf{10}$ & $\Lambda_1$ & 9 \\
($F$, $\bar{h}_{45}$) & ($T$, $H_{75}$) & $\mathbf{40}$ & $\Lambda_1$ & 0 \\
($F$, $H_{1}$) & ($T$, $\bar{h}_5$) & $\mathbf{5}$ & $\Lambda_1$ & 1  \\
($F$, $H_{1}$) & ($T$, $\bar{h}_{45}$) & $\mathbf{5}$ & $\Lambda_1$ & -3     \\
($F$, $H_{24}$) & ($T$, $\bar{h}_5$) & $\mathbf{5}$ & $\Lambda_1$ & -3/2 \\
($F$, $H_{24}$) & ($T$, $\bar{h}_5$) & $\mathbf{45}$ & $\Lambda_1$ & 3/2 \\
($F$, $H_{75}$) & ($T$, $\bar{h}_5$) & $\mathbf{45}$ & $\Lambda_1$ & -3 \\
($F$, $H_{24}$) & ($T$, $\bar{h}_{45}$) & $\mathbf{5}$ & $\Lambda_1$ & 9/2 \\
($F$, $H_{24}$) & ($T$, $\bar{h}_{45}$) & $\mathbf{45}$ & $\Lambda_1$ & -1/2 \\
($F$, $H_{75}$) & ($T$, $\bar{h}_{45}$) & $\mathbf{45}$ & $\Lambda_1$ & 1 \\
($F$, $H_{75}$) & ($T$, $\bar{h}_{45}$) & $\mathbf{50}$ & $\Lambda_1$ & 0 \\
\midrule
($F$, $\bar{h}_5$) & ($T$, $H_{1}$) & $\mathbf{10}$ & $\Lambda_{24}$ & 1/6 \\
($F$, $\bar{h}_{45}$) & ($T$, $H_{1}$) & $\mathbf{10}$ & $\Lambda_{24}$ & -1/2 \\
($F$, $\bar{h}_5$) & ($T$, $H_{24}$) & $\mathbf{10}$ & $\Lambda_{24}$ & 1 \\
($F$, $\bar{h}_5$) & ($T$, $H_{24}$) & $\mathbf{15}$ & $\Lambda_{24}$ & 0 \\
($F$, $\bar{h}_5$) & ($T$, $H_{75}$) & $\mathbf{10}$ & $\Lambda_{24}$ & -1/2   \\
($F$, $\bar{h}_{45}$) & ($T$, $H_{24}$) & $\mathbf{10}$ & $\Lambda_{24}$ & -3  \\
($F$, $\bar{h}_{45}$) & ($T$, $H_{24}$) & $\mathbf{40}$ & $\Lambda_{24}$ & 0   \\
($F$, $\bar{h}_{45}$) & ($T$, $H_{75}$) & $\mathbf{10}$ & $\Lambda_{24}$ & 3/2 \\
($F$, $\bar{h}_{45}$) & ($T$, $H_{75}$) & $\mathbf{40}$ & $\Lambda_{24}$ & 0   \\
($F$, $H_{1}$) & ($T$, $\bar{h}_5$) & $\mathbf{5}$ & $\Lambda_{24}$ & -2/3     \\
($F$, $H_{1}$) & ($T$, $\bar{h}_{45}$) & $\mathbf{5}$ & $\Lambda_{24}$ & 2     \\
($F$, $H_{24}$) & ($T$, $\bar{h}_5$) & $\mathbf{5}$ & $\Lambda_{24}$ & 1 \\
($F$, $H_{24}$) & ($T$, $\bar{h}_{45}$) & $\mathbf{5}$ & $\Lambda_{24}$ & -3 \\
($F$, $H_{75}$) & ($T$, $\bar{h}_{45}$) & $\mathbf{50}$ & $\Lambda_{24}$ & 0 \\
\bottomrule
\end{tabular}
\end{center}
\caption{
Resulting predictions for the SU(5) GUT scale Yukawa coupling ratios $(Y_e)_{ij}/(Y_d)_{ij}$
from the diagram in Fig.~\ref{Fig:Dim5}; for more details see main
text.
\label{Tab:SU5Messenger1}
}
\end{table}

In general, when $\Lambda$ is not a singlet but e.g.\ in the adjoint representation,
the components of the messenger pair $X$ and $\bar X$ receive masses (around
the GUT scale in size), which are split by CG coefficients. Below this scale,
since the GUT symmetry is now already spontaneously broken, also the fields in
the matter and Higgs representations are split into their SM representations
and, furthermore, the Higgs fields $B_1$ and $B_2$ acquire their vevs which is smaller
than the vev of $\Lambda$. One of the fields $B_1$ and $B_2$ contains the SM Higgs field and the other one
may either be a GUT singlet or in a 24- or 75-dimensional representation.
If  he field is not a GUT singlet, it introduces another CG factor. In total, we obtain
a prediction for the ratio $(Y_e)_{ij}/(Y_d)_{ij}$ of the considered charged lepton
and down-type quark Yukawa matrix element.

For illustration, let us study an instructive example: 
suppose $A = F$, $B_1 = H_1$, $B_2 = \bar h_5$
and $C=T$. $H_1$ is a GUT singlet acquiring a heavy vev satisfying $\langle H_1 \rangle < M_{\text{GUT}}$.
In a flavour model this could be for example a flavon. If the messenger
masses have a trivial mass proportional to $\langle \Lambda_1 \rangle$
we get the ordinary bottom-tau Yukawa coupling unification.
But we assume now that the symmetries are such that the pair $X$ and
$\bar X$, which are five-dimensional representations of SU(5), get their
mass from $\langle \Lambda_{24} \rangle$. To be more precise, when the mass
of the down-type quark like components of $X$ and $\bar X$
obtains a mass $- 2 \, M$, then the leptonic components
have the mass $3 \, M$ (with $M$ around $M_{\text{GUT}}$). 
At the GUT scale we therefore
get in the Lagrangian the effective operators
\begin{equation}
 \mathcal{L} \supset Y_{ij} \frac{\langle H_1 \rangle}{M} \left( \frac{1}{3} L_{i} \bar{e}_{j}^c - \frac{1}{2} Q_{j} \bar{d}_{i}^c   \right) h_d + \text{ H.c.,}
\end{equation}
which we have written down in terms of Standard Model fields
and where $i$, $j$ are family indices. Taking the ratio 
now yields $(Y_e)_{ij}/(Y_d)_{ij} = -2/3$, which is a new result not contained in \cite{Antusch:2009gu}. 

In Tab.~\ref{Tab:SU5Relations} we have collected all the ratios
we have found using this approach
where we have restricted $\Lambda$ to be not larger
than the adjoint and $X$ not to be larger than the $\mathbf{50}$ of SU(5).
The relations which are new compared to the
previous publication are highlighted in red. The involved
fields can be read off from Tab.~\ref{Tab:SU5Messenger1}
where we give the explicit representations for $A$,
$B_1$, $B_2$, $C$, $X$ and $\Lambda$ from
Fig.~\ref{Fig:Dim5} and their prediction for $(Y_e)_{ij}/(Y_d)_{ij}$.

There are a few more comments in order. First of all, in
Tab.~\ref{Tab:SU5Messenger1} we give only the cases
where the messenger fields connect the pairs $(A,B_1)$
and $(C,B_2)$ because only in this case the messengers
are matter-like and we get new relations. The other case
is given in the appendix, see especially Fig.~\ref{Fig:Other}.
In that case the messengers act as effective Higgs
fields which obtain an induced vev 
and we only get the well-known dimension four results for
the Yukawa couplings.

Then we also want to mention the case where $X$ and
$\bar X$ are not in conjugated representations to each
other. For example $X = \mathbf{5}$ and $\bar X =
\overline{\mathbf{45}}$ is possible if $\Lambda$ is the
adjoint of SU(5). In this case 40 of the 45 components
of $\bar X$ remain massless which should not be the case
to avoid light exotics. In principle one might add a $X' = \mathbf{45}$
field to overcome this problem but this might introduce
a different Yukawa coupling ratio and hence spoil the clean
prediction for $(Y_e)_{ij}/(Y_d)_{ij}$ so that we do not discuss
this case here any further.

Furthermore, it can happen that the product of $X$ with $\bar X$
contains more than one adjoint representation. To be more precise
this happens in SU(5) if the messengers are in 40- or 45-dimensional
representations and their products contain two different adjoint
representations. Evaluating the effective operator then might yield
two different ratios $(Y_e)_{ij}/(Y_d)_{ij}$ depending on how
you contract the indices.
For example, the combination $(A, B_1) = (F, H_{75})$ with $(C, B_2)=(T, \bar{h}_5)$
and $X = \mathbf{45}$ yields the two different ratios $(Y_e)_{ij}/(Y_d)_{ij} = 2$
and $(Y_e)_{ij}/(Y_d)_{ij} = 12$.
Hence, in these cases no proper prediction is possible
and we discard them from our tables.
Nevertheless, for the 40-dimensional messengers
the ratio does not depend on the SU(5) index
contraction so that the ratio remains unique and
we list it in the tables.
\footnote{We want to thank Vinzenz Maurer for bringing this
point to our attention.}

At this point we would like to make another remark.
There may of course be corrections to the predicted ratios,
coming for example from further operators at a higher level in the
operator expansion, from GUT threshold corrections or from the
effect of possible large GUT representations on the gauge
coupling evolution. The size of such corrections has to be
estimated when an explicit model is constructed, as was done,
for instance in an SO(10) model \cite{Aulakh:2008sn}.

\section{Predictions from Pati-Salam Unification}
\label{Sec:PS}

\begin{table}
\begin{center}
\begin{tabular}{cc} \toprule
Operator Dimension &	$((Y_e)_{ij}/(Y_d)_{ij} , (Y_u)_{ij}/(Y_d)_{ij}, (Y_\nu)_{ij}/(Y_u)_{ij} )$\\
\midrule 4	&	(1, 1, 1)     \\
	&	(-3, 1, -3)     \\
\midrule 5	&	{\color{red} (0, $\pm$1, 0)} \\
	&	{\color{red} (-1/3, $\pm$1, -1/3)} \\
	&	(1, $\pm$1, 1)     \\
	&	{\color{red} (3/2, $\pm$1, 3/2)} \\
	&	(-3, $\pm$1, -3)     \\
	&	(9, $\pm$1, 9)     \\
\bottomrule
\end{tabular}
\end{center}
\caption{
Summary of possible PS predictions for the GUT scale
Yukawa coupling ratios
$((Y_e)_{ij}/(Y_d)_{ij} , (Y_u)_{ij}/(Y_d)_{ij}, (Y_\nu)_{ij}/(Y_u)_{ij} )$.
The new relations compared
to \cite{Antusch:2009gu} are shown in red. For more details
about which operator gives which ratio, see Tabs.~\ref{Tab:PSMessenger1}
and \ref{Tab:PSMessenger2}. Partial results for dimension six
operators can be found in \protect\cite{Allanach:1996hz}
where $\Lambda$ was also taken to be a singlet.
 \label{Tab:PSRelations}
 }
\end{table}

\begin{table}
\begin{center}
\begin{tabular}{ccccc} \toprule
($A$, $B_1$) & ($C$, $B_2$) & $X$ & $\Lambda$ & $((Y_e)_{ij}/(Y_d)_{ij} , (Y_u)_{ij}/(Y_d)_{ij}, (Y_\nu)_{ij}/(Y_u)_{ij} )$ \\ \midrule
($R$, $h_{1}$) & ($\bar{R}$, $\phi^{+}_{1}$) & $(\mathbf{\bar{4}},\mathbf{1},\mathbf{\bar{2}})$
& $\Lambda^{\pm}_1$  & (1, $\pm$1, 1) \\
($R$, $h_{1}$) & ($\bar{R}$, $\phi^{+}_{1}$) & $(\mathbf{\bar{4}},\mathbf{1},\mathbf{\bar{2}})$
& $\Lambda^{\pm}_{15}$  & (-1/3, $\pm$1, -1/3) \\
($R$, $h_{1}$) & ($\bar{R}$, $\phi^{-}_{1}$) & $(\mathbf{\bar{4}},\mathbf{1},\mathbf{\bar{2}})$
& $\Lambda^{\pm}_1$  & (1, $\mp$1, 1) \\
($R$, $h_{1}$) & ($\bar{R}$, $\phi^{-}_{1}$) & $(\mathbf{\bar{4}},\mathbf{1},\mathbf{\bar{2}})$
& $\Lambda^{\pm}_{15}$  & (-1/3, $\mp$1, -1/3) \\
($R$, $h_{1}$) & ($\bar{R}$, $\phi^{+}_{15}$) & $(\mathbf{\bar{4}},\mathbf{1},\mathbf{\bar{2}})$
& $\Lambda^{\pm}_1$  & (-3, $\pm$1, -3) \\
($R$, $h_{1}$) & ($\bar{R}$, $\phi^{+}_{15}$) & $(\mathbf{\bar{4}},\mathbf{1},\mathbf{\bar{2}})$
& $\Lambda^{\pm}_{15}$  & (1, $\pm$1, 1) \\
($R$, $h_{1}$) & ($\bar{R}$, $\phi^{-}_{15}$) & $(\mathbf{\bar{4}},\mathbf{1},\mathbf{\bar{2}})$
& $\Lambda^{\pm}_1$  & (-3, $\mp$1, -3) \\
($R$, $h_{1}$) & ($\bar{R}$, $\phi^{-}_{15}$) & $(\mathbf{\bar{4}},\mathbf{1},\mathbf{\bar{2}})$
& $\Lambda^{\pm}_{15}$  & (1, $\mp$1, 1) \\
\midrule
($R$, $h_{15}$) & ($\bar{R}$, $\phi^{+}_{1}$) & $(\mathbf{\bar{4}},\mathbf{1},\mathbf{\bar{2}})$
& $\Lambda^{\pm}_1$  & (-3, $\pm$1, -3) \\
($R$, $h_{15}$) & ($\bar{R}$, $\phi^{+}_{1}$) & $(\mathbf{\bar{4}},\mathbf{1},\mathbf{\bar{2}})$
& $\Lambda^{\pm}_{15}$  & (1, $\pm$1, 1) \\
($R$, $h_{15}$) & ($\bar{R}$, $\phi^{-}_{1}$) & $(\mathbf{\bar{4}},\mathbf{1},\mathbf{\bar{2}})$
& $\Lambda^{\pm}_1$  & (-3, $\mp$1, -3) \\
($R$, $h_{15}$) & ($\bar{R}$, $\phi^{-}_{1}$) & $(\mathbf{\bar{4}},\mathbf{1},\mathbf{\bar{2}})$
& $\Lambda^{\pm}_{15}$  & (1, $\mp$1, 1) \\
($R$, $h_{15}$) & ($\bar{R}$, $\phi^{+}_{15}$) & $(\mathbf{\bar{4}},\mathbf{1},\mathbf{\bar{2}})$
& $\Lambda^{\pm}_1$  & (9, $\pm$1, 9) \\
($R$, $h_{15}$) & ($\bar{R}$, $\phi^{+}_{15}$) & $(\mathbf{\bar{4}},\mathbf{1},\mathbf{\bar{2}})$
& $\Lambda^{\pm}_{15}$  & (-3, $\pm$1, -3) \\
($R$, $h_{15}$) & ($\bar{R}$, $\phi^{-}_{15}$) & $(\mathbf{\bar{4}},\mathbf{1},\mathbf{\bar{2}})$
& $\Lambda^{\pm}_1$  & (9, $\mp$1, 9) \\
($R$, $h_{15}$) & ($\bar{R}$, $\phi^{-}_{15}$) & $(\mathbf{\bar{4}},\mathbf{1},\mathbf{\bar{2}})$
& $\Lambda^{\pm}_{15}$  & (-3, $\mp$1, -3) \\
\midrule
($R$, $h_{15}$) & ($\bar{R}$, $\phi^{+}_{15}$) & $(\mathbf{\overline{20}},\mathbf{1},\mathbf{\bar{2}})$
& $\Lambda^{\pm}_1$  & (0, $\pm$1, 0) \\
($R$, $h_{15}$) & ($\bar{R}$, $\phi^{+}_{15}$) & $(\mathbf{\overline{20}},\mathbf{1},\mathbf{\bar{2}})$
& $\Lambda^{\pm}_{15}$  & (0, $\pm$1,  0) \\
($R$, $h_{15}$) & ($\bar{R}$, $\phi^{-}_{15}$) & $(\mathbf{\overline{20}},\mathbf{1},\mathbf{\bar{2}})$
& $\Lambda^{\pm}_1$  & (0, $\mp$1, 0) \\
($R$, $h_{15}$) & ($\bar{R}$, $\phi^{-}_{15}$) & $(\mathbf{\overline{20}},\mathbf{1},\mathbf{\bar{2}})$
& $\Lambda^{\pm}_{15}$  & (0, $\mp$1, 0) \\
\midrule
($R$, $h_{15}$) & ($\bar{R}$, $\phi^{+}_{15}$) & $(\mathbf{\overline{36}},\mathbf{1},\mathbf{\bar{2}})$
& $\Lambda^{\pm}_1$  & (3/2, $\pm$1, 3/2) \\
($R$, $h_{15}$) & ($\bar{R}$, $\phi^{-}_{15}$) & $(\mathbf{\overline{36}},\mathbf{1},\mathbf{\bar{2}})$
& $\Lambda^{\pm}_1$  & (3/2, $\mp$1, 3/2) \\
\bottomrule
\end{tabular}
\end{center}
\caption{
Resulting predictions for the PS GUT scale Yukawa coupling ratios
$((Y_e)_{ij}/(Y_d)_{ij}$, $(Y_u)_{ij}/(Y_d)_{ij}$, $(Y_\nu)_{ij}/(Y_u)_{ij} )$
from the diagram in Fig.~\ref{Fig:Dim5}; for more details see main
text.
\label{Tab:PSMessenger1}}
\end{table}

\begin{table}
\begin{center}
\begin{tabular}{ccccc} \toprule
($A$, $B_1$) & ($C$, $B_2$) & $X$ & $\Lambda$ & $((Y_e)_{ij}/(Y_d)_{ij} , (Y_u)_{ij}/(Y_d)_{ij}, (Y_\nu)_{ij}/(Y_u)_{ij} )$ \\ \midrule
($R$, $\phi^{-}_{1}$) & ($\bar{R}$, $h_{1}$) & $(\mathbf{\bar{4}},\mathbf{\bar{2}},\mathbf{3})$
& $\Lambda^{+}_1$  & (1, -1, 1) \\
($R$, $\phi^{-}_{1}$) & ($\bar{R}$, $h_{1}$) & $(\mathbf{\bar{4}},\mathbf{\bar{2}},\mathbf{3})$
& $\Lambda^{+}_{15}$  & (-1/3, -1, -1/3) \\
($R$, $\phi^{-}_{1}$) & ($\bar{R}$, $h_{15}$) & $(\mathbf{\bar{4}},\mathbf{\bar{2}},\mathbf{3})$
& $\Lambda^{+}_1$  & (-3, -1, -3) \\
($R$, $\phi^{-}_{1}$) & ($\bar{R}$, $h_{15}$) & $(\mathbf{\bar{4}},\mathbf{\bar{2}},\mathbf{3})$
& $\Lambda^{+}_{15}$  & (1, -1, 1) \\
($R$, $\phi^{-}_{15}$) & ($\bar{R}$, $h_{1}$) & $(\mathbf{\bar{4}},\mathbf{\bar{2}},\mathbf{3})$
& $\Lambda^{+}_1$  & (-3, -1, -3) \\
($R$, $\phi^{-}_{15}$) & ($\bar{R}$, $h_{1}$) & $(\mathbf{\bar{4}},\mathbf{\bar{2}},\mathbf{3})$
& $\Lambda^{+}_{15}$  & (1, -1, 1) \\
($R$, $\phi^{-}_{15}$) & ($\bar{R}$, $h_{15}$) & $(\mathbf{\bar{4}},\mathbf{\bar{2}},\mathbf{3})$
& $\Lambda^{+}_1$  & (9, -1, 9) \\
($R$, $\phi^{-}_{15}$) & ($\bar{R}$, $h_{15}$) & $(\mathbf{\bar{4}},\mathbf{\bar{2}},\mathbf{3})$
& $\Lambda^{+}_{15}$  & (-3, -1, -3) \\
\midrule
($R$, $\phi^{+}_{1}$) & ($\bar{R}$, $h_{1}$) & $(\mathbf{\bar{4}},\mathbf{\bar{2}},\mathbf{1})$
& $\Lambda^{+}_1$  & (1, 1, 1) \\
($R$, $\phi^{+}_{1}$) & ($\bar{R}$, $h_{1}$) & $(\mathbf{\bar{4}},\mathbf{\bar{2}},\mathbf{1})$
& $\Lambda^{+}_{15}$  & (-1/3, 1, -1/3) \\
($R$, $\phi^{+}_{1}$) & ($\bar{R}$, $h_{15}$) & $(\mathbf{\bar{4}},\mathbf{\bar{2}},\mathbf{1})$
& $\Lambda^{+}_1$  & (-3, 1, -3) \\
($R$, $\phi^{+}_{1}$) & ($\bar{R}$, $h_{15}$) & $(\mathbf{\bar{4}},\mathbf{\bar{2}},\mathbf{1})$
& $\Lambda^{+}_{15}$  & (1, 1, 1) \\
($R$, $\phi^{+}_{15}$) & ($\bar{R}$, $h_{1}$) & $(\mathbf{\bar{4}},\mathbf{\bar{2}},\mathbf{1})$
& $\Lambda^{+}_1$  & (-3, 1, -3) \\
($R$, $\phi^{+}_{15}$) & ($\bar{R}$, $h_{1}$) & $(\mathbf{\bar{4}},\mathbf{\bar{2}},\mathbf{1})$
& $\Lambda^{+}_{15}$  & (1, 1, 1) \\
($R$, $\phi^{+}_{15}$) & ($\bar{R}$, $h_{15}$) & $(\mathbf{\bar{4}},\mathbf{\bar{2}},\mathbf{1})$
& $\Lambda^{+}_1$  & (9, 1, 9) \\
($R$, $\phi^{+}_{15}$) & ($\bar{R}$, $h_{15}$) & $(\mathbf{\bar{4}},\mathbf{\bar{2}},\mathbf{1})$
& $\Lambda^{+}_{15}$  & (-3, 1, -3) \\
\midrule
($R$, $\phi^{+}_{15}$) & ($\bar{R}$, $h_{15}$) & $(\mathbf{\overline{20}},\mathbf{\bar{2}},\mathbf{1})$
& $\Lambda^{+}_1$  & (0, 1, 0) \\
($R$, $\phi^{+}_{15}$) & ($\bar{R}$, $h_{15}$) & $(\mathbf{\overline{20}},\mathbf{\bar{2}},\mathbf{1})$
& $\Lambda^{+}_{15}$  & (0, 1, 0) \\
($R$, $\phi^{-}_{15}$) & ($\bar{R}$, $h_{15}$) & $(\mathbf{\overline{20}},\mathbf{\bar{2}},\mathbf{3})$
& $\Lambda^{+}_1$  & (0, -1, 0) \\
($R$, $\phi^{-}_{15}$) & ($\bar{R}$, $h_{15}$) & $(\mathbf{\overline{20}},\mathbf{\bar{2}},\mathbf{3})$
& $\Lambda^{+}_{15}$  & (0, -1, 0) \\
\midrule
($R$, $\phi^{+}_{15}$) & ($\bar{R}$, $h_{15}$) & $(\mathbf{\overline{36}},\mathbf{\bar{2}},\mathbf{1})$
& $\Lambda^{+}_1$  & (3/2, 1, 3/2) \\
($R$, $\phi^{-}_{15}$) & ($\bar{R}$, $h_{15}$) & $(\mathbf{\overline{36}},\mathbf{\bar{2}},\mathbf{3})$
& $\Lambda^{+}_1$  & (3/2, -1, 3/2) \\
\bottomrule
\end{tabular}
\end{center}
\caption{
Resulting predictions for the PS GUT scale Yukawa coupling ratios
$((Y_e)_{ij}/(Y_d)_{ij}$, $(Y_u)_{ij}/(Y_d)_{ij}$, $(Y_\nu)_{ij}/(Y_u)_{ij} )$
from the diagram in Fig.~\ref{Fig:Dim5}; for more details see main
text.
\label{Tab:PSMessenger2}}
\end{table}

For the case of PS models we followed in principle the same
approach as for SU(5), see previous section. Therefore we
will not describe the approach here again in detail but instead just
define all the fields and representations of the matter, Higgs
and $\Lambda$ fields involved.

The PS group SU(4)$_C \times $SU(2)$_L \times$SU(2)$_R$
is left-right symmetric and the matter fields of the Standard Model
are contained in two representations
\begin{align}
  R^{i}_{\alpha a} & =  (\mathbf{4},\mathbf{2},\mathbf{1})^i =
         \begin{pmatrix}
         u_L^R & u_L^B & u_L^G & \nu_L \\
         d_L^R & d_L^B & d_L^G & e_L^-
         \end{pmatrix}^i , \\
  \bar{R}^{i \alpha x} & =  (\mathbf{\overline{4}},\mathbf{1},\mathbf{\overline{2}})^i  =
        \begin{pmatrix}
         \bar{d}_R^R & \bar{d}_R^B & \bar{d}_R^G & e_R^+ \\
         \bar{u}_R^R & \bar{u}_R^B & \bar{u}_R^G & \bar{\nu}_R
         \end{pmatrix}^i ,
\end{align} 
where $\alpha = 1,\ldots,4$ is an SU(4)$_C$ index,
$a,x = 1,2$ are SU(2)$_{L,R}$ indices and $i=1,2,3$ is
a family index. The fields in $R^i$ form SU(2)$_L$
doublets and the fields in $\bar{R}^i$ 
SU(2)$_R$ doublets as indicated by the indices $L$ and $R$.

The MSSM Higgs doublets are contained in the bi-doublet representation
\begin{equation}
  (h_1)_x^a = (\mathbf{1},\overline{\mathbf{2}},\mathbf{2}) = \begin{pmatrix}
         h_u^+ & h_d^0 \\
         h_u^0 & h_d^-
         \end{pmatrix} \;,
\end{equation}
where the components $h_u^0$ and $h_d^0$ acquire the electroweak
symmetry breaking vevs. To get the Georgi-Jarlskog relation at the
renormalisable level the Higgs doublets are contained in the
$h_{15} = (\mathbf{15}, \mathbf{\bar{2}}, \mathbf{2})$ field where
the vev points into the direction of $B-L$ due to the tracelessness
of the adjoint of SU(4)$_C$. 

The PS symmetry is broken by the two Higgs fields
\begin{align}
  H^{\alpha b} &= (\mathbf{4},\mathbf{1},\mathbf{2}) = \begin{pmatrix}
         u_H^R & u_H^B & u_H^G & \nu_H \\
         d_H^R & d_H^B & d_H^G & e_H^-
         \end{pmatrix} \;, \\
  \bar{H}_{\alpha x} &= (\bar{\mathbf{4}},\mathbf{1},\bar{\mathbf{2}}) = \begin{pmatrix}
         \bar{d}_H^R & \bar{d}_H^B & \bar{d}_H^G & e_H^+ \\
         \bar{u}_H^R & \bar{u}_H^B & \bar{u}_H^G & \bar{\nu}_H
         \end{pmatrix} \;,
\end{align}
where the GUT symmetry breaking vev points in the directions
$\langle {\nu}_H \rangle$ and $\langle {\bar{\nu}}_H \rangle$.
We have also considered the cases where adjoints of PS
acquire GUT scale vevs
\begin{align}
 \phi^{+}_{1} &= (\mathbf{1},\mathbf{1},\mathbf{1}) \;, \quad  \phi^{+}_{15} = (\mathbf{15},\mathbf{1},\mathbf{1}) \;, \\
 \phi^{-}_{1} &= (\mathbf{1},\mathbf{1},\mathbf{3}) \;, \quad  \phi^{-}_{15} = (\mathbf{15},\mathbf{1},\mathbf{3})
 \;,
\end{align}
which would not break PS to the Standard Model. 
However, in a more complete theory, one could regard them as effective combinations of $H \bar H$,
as discussed in \cite{Allanach:1997gu}.

In the PS case we have more possibilities for the $\Lambda$
fields because we can take adjoints of SU(4)$_C$ and SU(2)$_R$
and combinations of them
\begin{align}
 \Lambda^{+}_{1} &= (\mathbf{1},\mathbf{1},\mathbf{1}) \;, \quad  \Lambda^{+}_{15} = (\mathbf{15},\mathbf{1},\mathbf{1}) \;, \\
 \Lambda^{-}_{1} &= (\mathbf{1},\mathbf{1},\mathbf{3}) \;, \quad  \Lambda^{-}_{15} = (\mathbf{15},\mathbf{1},\mathbf{3})
 \;.
\end{align}

The results for PS are summarized in Tab.~\ref{Tab:PSRelations} and the
detailed operators are listed in Tabs.~\ref{Tab:PSMessenger1} and
\ref{Tab:PSMessenger2}. Also for the PS case we find new relations
compared to the previous study \cite{Antusch:2009gu}, for example
we find the ratio $(Y_e)_{ij}/(Y_d)_{ij} = 3/2$ which is a promising
ratio for the third generation as it was noted in \cite{Antusch:2009gu}
but where (with a singlet $\Lambda$) this ratio only appeared in the SU(5) case.

We note that there are certain possible combinations
of external fields which are not contained in the tables. There are four
possible reasons for this.
The first reason is that we have put into the appendix again the cases where
the messenger fields are Higgs-like and are thus not giving any new results
beyond the renormalisable dimension four operators. The second reason is that,
as in SU(5), the fields $X$ and $\bar X$ are not conjugated to each other
introducing an extra model building complication which we do not want to discuss here.
The third reason also appeared already in SU(5), namely the case when the product
of the messengers contains more than one adjoint representation. In PS this happens
for the 15-, 20- and 36-dimensional representation of SU(4)$_C$. We have again dropped
the non-unique predictions which affected here only the 36-dimensional representations.

Apart from these three reasons, in the case of PS models another reason has
emerged from our analysis, namely that it can happen that even if $X$ and $\bar X$ 
are in conjugated representations to each other, 
some messenger components remain massless if $\Lambda$ is a non-singlet.
For the SU(2)$_R$ part this can be understood easily as follows: 
Remembering that the adjoint is a traceless tensor we have
\begin{equation}
 \langle \Lambda^{-}_1 \rangle = \Lambda \begin{pmatrix}
  1 & 0 \\ 0 & -1
\end{pmatrix} \;.
\end{equation}
Suppose now $X$ and $\bar X$ are adjoints of SU(2)$_R$ as well
with
\begin{equation}
 X = \begin{pmatrix}
  X^0 & X^+ \\ X^- & -X^0
\end{pmatrix} \;,
\end{equation}
and similar for $\bar X$
The mass term generated by the vev of $\Lambda^{-}_1$ is then
\begin{equation}
 \mathcal{W} \supset \text{Tr}(X \, \langle \Lambda^{-}_1 \rangle \,  \bar X) = \Lambda ( X^+ \bar X^- + X^- \bar X^+ ) \;,
\end{equation}
which implies that there is no mass term for $X^0$ and $\bar X^0$.
Similar to the case where $X$ and $\bar X$ are not conjugated to each
other this deficit might be fixed by introducing an extra field (in this case
another $\Lambda$ field) which might however spoil the clean
prediction so that we decided not to list such cases as well.

\section{Generalisation to higher order operators}

\begin{figure}
\centering
\includegraphics[scale=0.7,angle=270]{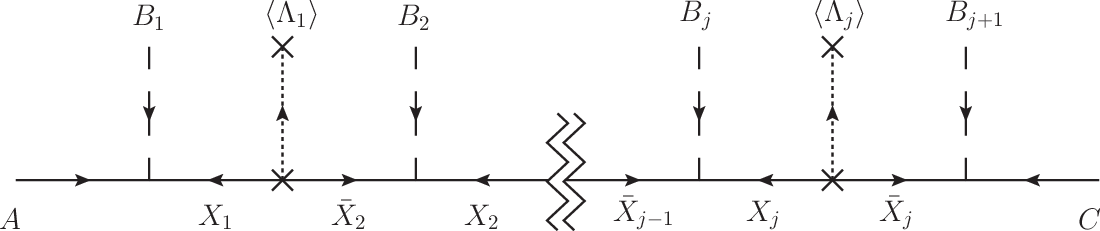}
\caption{
Generalisation of Fig.~\ref{Fig:Dim5} with $j$ messenger pairs,
$j$ insertions of $\Lambda $ vevs and $j+1$ external $B$ fields
from which one receives an electroweak vev and $j$ receive high
scale vevs.
\label{Fig:General}
}
\end{figure}

So far we have discussed only relations coming from diagrams with only one
messenger pair $X$ and $\bar X$. Nevertheless, these results
can be generalised to diagrams with additional external heavy
fields and additional insertion of (non-trivial) messenger masses. Indeed, in
such a way it is possible to realise new CG factors, which can be
understood as products of the CG factors appearing at lower orders.
There are basically three effects which have an influence on the Yukawa coupling
ratio predicted by the respective diagram, as expounded in the following bullet points.

\begin{itemize}

\item Already at the renormalizable level the Yukawa coupling ratio depends on which 
representations contain the electroweak symmetry breaking Higgs doublets. In
SU(5) these are $\bar h_5$ and $\bar h_{45}$ (and $h_5$ and $h_{45}$ which do not
matter here because the up-type quark Yukawa couplings are not related to any other
Yukawa couplings in SU(5)). And in PS these are the fields $h_1$ and $h_{15}$. While
$\bar h_5$ and $h_1$ give unification of Yukawa couplings, $\bar h_{45}$ and $h_{15}$
give a relative factor of -3 between leptons and quarks (the vev points in the
direction of $B-L$). 

\item The second effect is associated with external fields receiving a GUT scale vev.
Let us consider for example the SU(5) diagram from Fig.~\ref{Fig:Dim5} with
$(A, B_1) = (F, H_{24})$ with $(C, B_2) = (T, \bar h_5)$ and $X = \mathbf{5}$. Then
$(Y_e)_{ij}/(Y_d)_{ij} = -3/2$ which is nothing else than the ratio of hypercharges of the fields
contained in the $\mathbf{5}$ of SU(5) because the vev of $H_{24}$ points into the
hypercharge direction. Looking at Fig.~\ref{Fig:General} it is straightforward to
see how this result generalises: inserting an additional
$H_{24}$ (as a $B_j$ in Fig.~\ref{Fig:General}) coupling to fiveplet messenger fields
contributes a factor of $-3/2$ to the resulting CG factor. The analogous consideration
can be done for an additional $H_{24}$ coupling to tenplet messenger fields. This yields an 
additional factor of $6$. For the PS case similar arguments apply. 

\item Thirdly, the resulting CG factor can be affected when the messenger masses arise
from vevs of $\Lambda_i$ fields which are not gauge singlets. This leads to split masses
for the component fields of the messengers, and thus to inverse CG factors as we
discussed in the previous sections. As shown in Fig.~\ref{Fig:General}, also this
mechanism can be generalised. In the SU(5) case,
this generalisation is particularly simple because the mass terms we consider are
either universal or coming from an adjoint giving the inverse of the hypercharge ratio
for the Yukawa coupling ratio. For each non-trivial messenger mass from an adjoint, we
therefore obtain a factor of $(-3/2)^{-1}$ for five-dimensional messengers and a factor
of $(1/6)^{-1}$ for ten-dimensional messengers. Again, similar considerations can be done
for the PS case.
\end{itemize}

\begin{figure}
\centering
\includegraphics[scale=0.7]{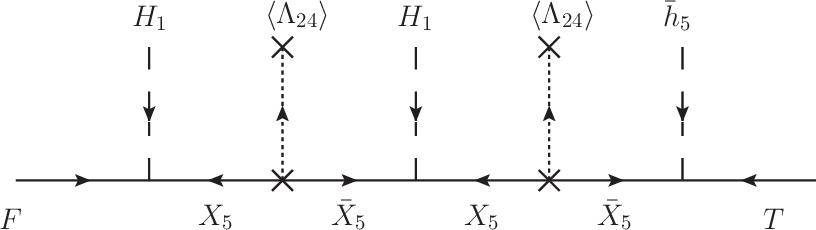}
\caption{
Example for a generalisation of Fig.~\ref{Fig:Dim5}
(of the form shown in Fig.~\ref{Fig:General}) which leads to a CG factor of $4/9$. The messenger pairs $X_5$
and $\bar X_5$ are five-dimensional representations of SU(5).
The notation for the other fields and further details can be found in the main text. A possible application is discussed in section 5.
\label{Fig:4O9}
}
\end{figure}

Based on these considerations, one can construct new diagrams which
effectively generate products of CG factors. 
Instead of going through all possibilities at higher order,
let us consider an explicit example (which we then also apply in the next section).
Fig.~\ref{Fig:4O9} shows a diagram which illustrates how a CG factor of $4/9$ can be realised.
The two split messenger masses from the vevs of adjoint representations each contribute
a factor of $(-3/2)^{-1}$, such that the resulting CG factor is  $(-3/2)^{-2} = 4/9$.

\section{Applications}

The novel CG factors found in the previous sections open up new possibilities for
GUT model building. In this section, we will discuss some of these possibilities.
In particular, the new CG factors allow for alternative textures for the GUT-scale
Yukawa matrices, especially regarding the first two families. Let us briefly discuss
the situation in SU(5) as an example (following \cite{Antusch:2011qg,Marzocca:2011dh}):

In SU(5) GUTs, $Y_e$ is related to $Y_d^T$, and without loss of generality we can
write for the upper 1-2 block describing the first two families in the $2\times 2$ Yukawa matrices
\be
Y_d = \begin{pmatrix} 
d&b\\
a&c
\end{pmatrix}\:
\Rightarrow \quad
Y_e = 
\begin{pmatrix} 
c_d \, d& c_b \, b\\
 c_a\, a&c_c \, c
\end{pmatrix}^T
=
\begin{pmatrix} 
c_d \, d& c_a \, a\\
 c_b\, b&c_c \, c
\end{pmatrix},
\ee 
where we have introduced the CG  factors $c_a,c_b,c_c$ and $c_d$ for the
respective matrix elements. We assume here that each of the matrix elements arises
dominantly from one GUT operator and that the 1-3 and 2-3 mixing effects in
$Y_d$ and $Y_e$ can be neglected for discussing the mass relations for the first
two families. Possible GUT operators and their predicted CG factors have been
discussed in the previous sections. With $a,b,c$ and $d$ all being non-zero complex
numbers, this obviously results in a large number of possibilities. 

Many of these possibilities are however constrained by phenomenology. For instance
in the context of SUSY GUTs, to check the validity of such a texture, one has to
compute the RG evolution of the mass eigenvalues from the GUT scale to low energies
(or vice versa), taking into account the radiative threshold effects at the SUSY
scale. In a recent study \cite{Antusch:2013jca}, the following constraints for the
GUT scale values of the diagonal Yukawa couplings have been derived:
 \be
 \frac{(1 + \bar\eta_\ell) y_\mu}{(1 + \bar\eta_q) y_s}\approx 4.36 \pm 0.23\; , \quad 
 \frac{(1 + \bar\eta_\ell) y_e}{(1 + \bar\eta_q) y_d} \approx 0.41^{+0.02}_{-0.06} \: , \ee
where the $\bar \eta_i$ are threshold correction parameters and where the ranges indicate the $1\sigma$ uncertainties. 
The $\bar \eta_i$ enter both relations and they drop out in the following constraint equation for the
effective CG factors $c_e = y_e/y_d$ and $c_\mu =  y_\mu/y_s $ where $y_e$,
$y_\mu$, $y_d$ and $y_s$ are the Yukawa couplings of the electron, the muon,
the down quark and the strange quark \cite{Antusch:2013jca}:
\be
 \frac{y_\mu}{y_s} \left( \frac{y_e}{y_d} \right)^{-1} = \frac{c_\mu}{c_e} \approx 10.7^{+1.8}_{-0.8} \;  .
\ee
From this constraint, one can easily check the validity of the possible GUT textures
for the masses of the first two families. 

Another interesting consequence of these GUT textures is that, when embedded into a full
flavour GUT model, they give different 1-2 mixing contributions in the charged lepton sector,
as discussed systematically for the CG factors of \cite{Antusch:2009gu} in
\cite{Antusch:2011qg,Marzocca:2011dh}. In small mixing approximation, it is given as
$\theta_{12}^\mathrm{e} \approx \frac{c_a \, a}{c_c \, c}$. Its value (and phase) is
important for selecting possibly viable textures of the neutrino mass matrix and for
calculating the predictions for the leptonic mixing angles. For example, in combination
with a neutrino mass matrix with zero 1-3 mixing and negligible 1-3 mixing in $Y_e$, the
size of $\theta_{12}^\mathrm{e}$ controls the leptonic 1-3 mixing angle as
$\theta_{13}^\mathrm{PMNS} \approx  \sin(\theta_{23}^\mathrm{PMNS})\theta_{12}^\mathrm{e} \approx \theta_{12}^\mathrm{e} / \sqrt{2}$.
For possible choices of CG factors, which relate $\theta_{12}^\mathrm{e}$ to the Cabibbo angle $\theta_C$, the resulting predictions for $\theta_{13}^\mathrm{PMNS}$ have been studied in 
\cite{Antusch:2011qg,Marzocca:2011dh}. The specific relation $\theta_{13}^\mathrm{PMNS} \approx \theta_C/\sqrt{2}$, which 
is close to the measured value, has been discussed recently in the context of GUTs in \cite{King:2012vj,Antusch:2012fb}. 

\subsubsection*{Example 1: Alternative texures with zero $\boldsymbol{(Y_e)_{11}}$ and $\boldsymbol{(Y_d)_{11}}$ }

With $a,b,c,$ and $d$ all non-zero complex numbers, there are in general no predictions for the
quark-lepton mass ratios. Probably most popular predictive texture in SU(5) GUT model
building uses the Georgi-Jarlskog CG factor $c_c = 3$ and $c_a = c_b = 1$ while $d=0$, i.e.\
\be
Y_d = \begin{pmatrix} 
0&b\\
a&c
\end{pmatrix}\:
\Rightarrow \quad
Y_e = 
\begin{pmatrix} 
0 & b\\
 a&3 \, c
\end{pmatrix}^T
=
\begin{pmatrix} 
0&  a\\
  b&3 \, c
\end{pmatrix},
\ee 
often combined with the assumption of a symmetric matrix. This texture results (in LO small mixing approximation) in 
diagonal Yukawa couplings $y_d\approx ab/c$, $y_s\approx c$, $y_e\approx ab/(3c)$, $y_{\mu}\approx 3c$ and
hence ratios of diagonal Yukawa couplings $c_e = \frac{y_e}{y_d}$ and $c_\mu =  \frac{y_\mu}{y_s} $ given by,
\be
c_\mu \approx c_c = 3 \; , \quad c_e \approx \frac{c_a c_b}{c_c} = \frac{1}{3}\quad \Rightarrow
\quad \frac{c_\mu}{c_e} \approx 9   \;.
\ee
In addition, assuming $|\theta_{12}^\mathrm{d}| \approx \theta_C$ (which implies $a \approx b$), it leads to a small
charged lepton mixing contribution of about $\theta_{12}^\mathrm{e} \approx \frac{c_a \, a}{c_c \, c} \approx \theta_C /3$.

An alternative texture, with $|\theta_{12}^\mathrm{e}| \approx \theta_C$, and better agreement with the
experimental data, was highlighted in \cite{Antusch:2012fb}:
\be
Y_d = \begin{pmatrix} 
0&b\\
a&c
\end{pmatrix}\:
\Rightarrow \quad
Y_e = 
\begin{pmatrix} 
0 & \tfrac{1}{2}  b\\
 6\,a&6 \, c
\end{pmatrix}^T
=
\begin{pmatrix} 
0&  6\,a\\
\tfrac{1}{2}  b&6 \, c
\end{pmatrix},
\ee 
which implies
diagonal Yukawa couplings $y_d\approx ab/c$, $y_s\approx c$, $y_e\approx ab/(2c)$, $y_{\mu}\approx 6c$ and
hence ratios of diagonal Yukawa couplings $c_e = \frac{y_e}{y_d}$ and $c_\mu =  \frac{y_\mu}{y_s} $ given by,
\be
c_\mu \approx c_c = 6 \; , \quad c_e \approx \frac{c_a c_b}{c_c} = \frac{1}{2}\quad \Rightarrow
\quad \frac{c_\mu}{c_e} \approx 12   \;.
\ee
In this texture $c_a = c_c$ and thus (taking $|\theta_{12}^\mathrm{d}| \approx \theta_C$  which implies
$a \approx b$) one obtains $|\theta_{12}^\mathrm{e}| \approx \theta_C$. 
The texture has been applied recently to construct predictive flavour GUT models in
\cite{Antusch:2013kna,Antusch:2013tta}. In Tab.~\ref{Tab:SU5Messenger1} we have presented a
new way to obtain the CG factor $\tfrac{1}{2}$ using an ``inverse CG factor''.

With the new possible CG factor of $\tfrac{4}{9}$ from a higher-dimensional operator 
as discussed in section 4 there is another potentially interesting option
which is in good agreement with the data, namely
\be
Y_d = \begin{pmatrix} 
0&b\\
a&c
\end{pmatrix}\:
\Rightarrow \quad
Y_e = 
\begin{pmatrix} 
0 & \tfrac{4}{9}  b\\
 \tfrac{9}{2}  a&\tfrac{9}{2}   \, c   \vphantom{\frac{f}{f}}
\end{pmatrix}^T
=
\begin{pmatrix} 
0&  \tfrac{9}{2} a\\
\tfrac{4}{9}  b&\tfrac{9}{2}  \, c      \vphantom{\frac{f}{f}}
\end{pmatrix}.
\ee 
It implies
diagonal Yukawa couplings $y_d\approx ab/c$, $y_s\approx c$, $y_e\approx 4ab/(9c)$, $y_{\mu}\approx 9c/2$
and hence ratios of diagonal Yukawa couplings $c_e = \frac{y_e}{y_d}$ and $c_\mu =  \frac{y_\mu}{y_s} $ given by,
\be
c_\mu \approx c_c = \frac{9}{2} \; , \quad c_e \approx \frac{c_a c_b}{c_c} = \frac{4}{9} \quad \Rightarrow
\quad \frac{c_\mu}{c_e} \approx 10    \;.
\ee

\subsubsection*{Example 2: Alternative texures with diagonal $\boldsymbol{Y_e}$ and $\boldsymbol{Y_d}$}

Another highly predictive situation is the case that $Y_e$ and $Y_d$ are both diagonal, i.e. $a=b=0$ in the above notation,
\be
Y_d = \begin{pmatrix} 
d&0\\
0&c
\end{pmatrix}= \begin{pmatrix} 
y_d&0\\
0&y_s
\end{pmatrix}\:
\Rightarrow \quad
Y_e = 
\begin{pmatrix} 
c_d \, d&0 \\
0&c_c \, c
\end{pmatrix}
= \begin{pmatrix} 
y_e&0\\
0&y_{\mu}
\end{pmatrix}.
\ee 
Then, the ratios of diagonal Yukawa couplings
$c_e = \frac{y_e}{y_d}$ and $c_\mu =  \frac{y_\mu}{y_s} $
are simply given by
\be
c_e = c_d \; , \quad c_\mu = c_c \quad \Rightarrow \quad \frac{c_\mu}{c_e} =   \frac{c_c}{c_d}  \;. 
\ee

The new CG factor of $\tfrac{1}{3}$, available in Pati-Salam models,
can be used in the combination $c_c = 3$ and $c_d = \tfrac{1}{3}$ to obtain the same prediction for
the mass relations as from the Georgi-Jarlskog texture, with (as above) $\frac{c_\mu}{c_e} \approx 9$.
In contrast to the Georgi-Jarlskog texture it yields no charged lepton 1-2 mixing,
$\theta_{12}^\mathrm{d} \approx 0$, which is interesting in the context of neutrino mass textures
which generate all lepton mixing (including $\theta_{13}^\mathrm{PMNS}$) already in the neutrino
sector, e.g.\ \cite{King:2013iva,King:2013xba}. 

A viable option for SU(5) is the combination of CG factors $c_c = 6$ and $c_d = \tfrac{1}{2}$.
It yields: 
\be
c_\mu = c_c = 6 \; , \quad c_e = c_d = \frac{1}{2}\quad \Rightarrow
\quad \frac{c_\mu}{c_e} = 12.0   \;.
\ee
In Tab.~\ref{Tab:SU5Messenger1}
we have presented a new way to obtain the CG factor $\tfrac{1}{2}$ using an
``inverse CG factor''.

The CG factor $ \frac{4}{9}$ from a higher-dimensional operator (cf.\ section 4) can also be
used for the first family with diagonal $Y_e$ and $Y_d$, such that $c_e = c_d =  \frac{4}{9}$
directly. In combination with $c_c = \frac{9}{2}$, we obtain  
\be
c_\mu = c_c = \frac{9}{2} \; , \quad c_e = c_d = \frac{4}{9} \quad \Rightarrow \quad \frac{c_\mu}{c_e} = 10.1\;,
\ee
in good agreement with experiments.

\section{Conclusions}

We have proposed new GUT predictions for the ratios of quark and lepton Yukawa couplings arising
from splitting the masses of the messenger fields for the GUT scale Yukawa operators by CG factors
from GUT symmetry breaking. The effect is that the CG factors enter inversely in the predicted quark-lepton mass relations.
This allows new fractional CG factors for the ratios of charged lepton to down-type quark Yukawa couplings
such as $\tfrac{1}{6}$, -$\tfrac{2}{3}$ in SU(5) or -$\frac{1}{3}$, $\tfrac{3}{2}$ in Pati-Salam,
leading to new possible GUT predictions. 

We have systematically constructed the new predictions that can be realised this way in SU(5)
GUTs and Pati-Salam unified theories. The new predictions all arise from the types of diagrams in
Fig.~\ref{Fig:Dim5} and their generalisation to higher orders in Fig.~\ref{Fig:General}. In other words,
diagrams in which the messenger fields are matter-like and receive
their masses from non-singlet GUT representations getting vevs. 
The resulting possible new predictions are indicated in red in Tab.~\ref{Tab:SU5Relations} and 
Tab.~\ref{Tab:PSRelations}. 

Note that the diagrams of the type shown in Fig.~\ref{Fig:Other}, where the messengers are Higgs-like fields, 
always lead to the usual CG relations, even when the messenger fields receive masses from
non-singlet GUT representations getting vevs. This is because in these types of diagrams, one may think of all
such diagrams as giving an effective vev for the Higgs coupling directly to the quarks and leptons.

We have also discussed some possible new model building applications involving the new CG coefficients
in the case of both SU(5) GUTs and Pati-Salam unified theories. 
For example the new fractional CG coefficient of $\tfrac{4}{9}$ opens up some new interesting possibilities
for SU(5) and the coefficient of -$\tfrac{1}{3}$ for Pati-Salam unified theories.

\section*{Acknowledgements}
We thank Vinzenz Maurer for useful discussions. 
S.A.\ acknowledges support by the Swiss National Science Foundation,
S.F.K. from the STFC Consolidated ST/J000396/1 and
M.S.\ partially by the ERC Advanced Grant no. 267985 ``DaMESyFla''.
S.F.K. and M.S. also acknowledge partial support from the EU Marie Curie
ITN ``UNILHC'' (PITN-GA-2009-237920) and all authors were partially supported by the
European Union under FP7 ITN INVISIBLES (Marie Curie Actions, PITN-GA-2011-289442).

\appendix

\section{Cases with ``Higgs-like'' messengers}

\begin{figure}
\centering
\includegraphics[scale=0.8]{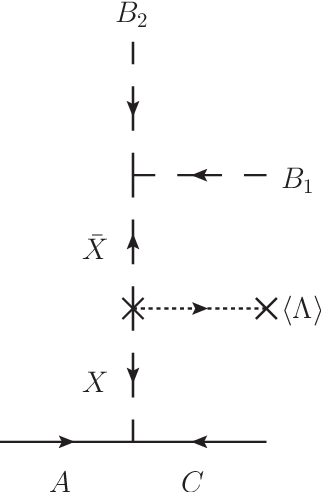}
\caption{
Variation of Fig.~\ref{Fig:Dim5} with a different topology
where the messenger fields are ``Higgs-like''.
\label{Fig:Other}
}
\end{figure}

\begin{table}
\begin{center}
\begin{tabular}{ccccc} \toprule
($A$, $C$) & ($B_1$, $B_2$) & $X$ & $\Lambda$ & $(Y_e)_{ij}/(Y_d)_{ij}$ \\ \midrule
($F$, $T$) & ($\bar{h}_5$, $H_{24}$) & $\overline{\mathbf{5}}$ & $\Lambda_1$ & 1 \\
($F$, $T$) & ($\bar{h}_5$, $H_{24}$) & $\overline{\mathbf{45}}$ & $\Lambda_1$ & -3 \\
($F$, $T$) & ($\bar{h}_5$, $H_{75}$) & $\overline{\mathbf{45}}$ & $\Lambda_1$ & -3 \\
($F$, $T$) & ($\bar{h}_{45}$, $H_{24}$) & $\overline{\mathbf{5}}$ & $\Lambda_1$ & 1 \\
($F$, $T$) & ($\bar{h}_{45}$, $H_{24}$) & $\overline{\mathbf{45}}_1$ & $\Lambda_1$ & -3 \\
($F$, $T$) & ($\bar{h}_{45}$, $H_{24}$) & $\overline{\mathbf{45}}_2$ & $\Lambda_1$ & -3 \\
($F$, $T$) & ($\bar{h}_{45}$, $H_{75}$) & $\overline{\mathbf{5}}$ & $\Lambda_1$ & 1 \\
($F$, $T$) & ($\bar{h}_{45}$, $H_{75}$) & $\overline{\mathbf{45}}_1$ & $\Lambda_1$ & -3 \\
($F$, $T$) & ($\bar{h}_{45}$, $H_{75}$) & $\overline{\mathbf{45}}_2$ & $\Lambda_1$ & -3 \\
\midrule
($F$, $T$) & ($\bar{h}_5$, $H_{1}$) & $\overline{\mathbf{5}}$ & $\Lambda_{24}$ & 1 \\
($F$, $T$) & ($\bar{h}_{45}$, $H_{1}$) & $\overline{\mathbf{45}}$ & $\Lambda_{24}$ & -3 \\
($F$, $T$) & ($\bar{h}_5$, $H_{24}$) & $\overline{\mathbf{5}}$ & $\Lambda_{24}$ & 1 \\
($F$, $T$) & ($\bar{h}_5$, $H_{24}$) & $\overline{\mathbf{45}}$ & $\Lambda_{24}$ & -3 \\
($F$, $T$) & ($\bar{h}_5$, $H_{75}$) & $\overline{\mathbf{45}}$ & $\Lambda_{24}$ & -3 \\
($F$, $T$) & ($\bar{h}_{45}$, $H_{24}$) & $\overline{\mathbf{5}}$ & $\Lambda_{24}$ & 1 \\
($F$, $T$) & ($\bar{h}_{45}$, $H_{24}$) & $\overline{\mathbf{45}}_1$ & $\Lambda_{24}$ & -3 \\
($F$, $T$) & ($\bar{h}_{45}$, $H_{24}$) & $\overline{\mathbf{45}}_2$ & $\Lambda_{24}$ & -3 \\
($F$, $T$) & ($\bar{h}_{45}$, $H_{75}$) & $\overline{\mathbf{5}}$ & $\Lambda_{24}$ & 1 \\
($F$, $T$) & ($\bar{h}_{45}$, $H_{75}$) & $\overline{\mathbf{45}}_1$ & $\Lambda_{24}$ & -3 \\
($F$, $T$) & ($\bar{h}_{45}$, $H_{75}$) & $\overline{\mathbf{45}}_2$ & $\Lambda_{24}$ & -3 \\
\bottomrule
\end{tabular}
\end{center}
\caption{
Resulting predictions for the SU(5) GUT scale Yukawa coupling ratios $(Y_e)_{ij}/(Y_d)_{ij}$
from the diagram in Fig.~\ref{Fig:Other}; for more details see main
text.
If the messenger representation $X$ has an index, there is more than one
way to combine the fields $A$ and $C$ or $B_1$ and $B_2$ to form this
representation.
\label{Tab:SU5Messenger2}
}
\end{table}

\begin{table}
\begin{center}
\begin{tabular}{ccccc} \toprule
($A$, $C$) & ($B_1$, $B_2$) & $X$ & $\Lambda$ & $((Y_e)_{ij}/(Y_d)_{ij}$, $(Y_u)_{ij}/(Y_d)_{ij}$, $(Y_\nu)_{ij}/(Y_u)_{ij} )$ \\ \midrule
($R$, $\bar{R}$) & ($h_{1}$, $\phi^{+}_{1}$) & $(\mathbf{1},\mathbf{\overline{2}},\mathbf{2})$
& $\Lambda^{\pm}_1$ & (1, $\pm$1, 1) \\
($R$, $\bar{R}$) & ($h_{1}$, $\phi^{-}_{1}$) & $(\mathbf{1},\mathbf{\overline{2}},\mathbf{2})$
& $\Lambda^{\pm}_1$ & (1, $\mp$1, 1) \\
($R$, $\bar{R}$) & ($h_{15}$, $\phi^{+}_{15}$) & $(\mathbf{1},\mathbf{\overline{2}},\mathbf{2})$
& $\Lambda^{\pm}_1$ & (1, $\pm$1, 1) \\
($R$, $\bar{R}$) & ($h_{15}$, $\phi^{-}_{15}$) & $(\mathbf{1},\mathbf{\overline{2}},\mathbf{2})$
& $\Lambda^{\pm}_1$ & (1, $\mp$1, 1) \\
\midrule
($R$, $\bar{R}$) & ($h_{15}$, $\phi^{+}_{1}$) & $(\mathbf{15},\mathbf{\overline{2}},\mathbf{2})$
& $\Lambda^{\pm}_1$  & (-3, $\pm$1, -3) \\
($R$, $\bar{R}$) & ($h_{1}$, $\phi^{+}_{15}$) & $(\mathbf{15},\mathbf{\overline{2}},\mathbf{2})$
& $\Lambda^{\pm}_1$  & (-3, $\pm$1, -3) \\
($R$, $\bar{R}$) & ($h_{1}$, $\phi^{-}_{15}$) & $(\mathbf{15},\mathbf{\overline{2}},\mathbf{2})$
& $\Lambda^{\pm}_1$  & (-3, $\mp$1, -3) \\
($R$, $\bar{R}$) & ($h_{15}$, $\phi^{-}_{1}$) & $(\mathbf{15},\mathbf{\overline{2}},\mathbf{2})$
& $\Lambda^{\pm}_1$  & (-3, $\mp$1, -3) \\
($R$, $\bar{R}$) & ($h_{15}$, $\phi^{+}_{15}$) & $(\mathbf{15}_1,\mathbf{\overline{2}},\mathbf{2})$
& $\Lambda^{\pm}_1$  & (-3, $\pm$1, -3) \\
($R$, $\bar{R}$) & ($h_{15}$, $\phi^{+}_{15}$) & $(\mathbf{15}_2,\mathbf{\overline{2}},\mathbf{2})$
& $\Lambda^{\pm}_1$  & (-3, $\pm$1, -3) \\
($R$, $\bar{R}$) & ($h_{15}$, $\phi^{-}_{15}$) & $(\mathbf{15}_1,\mathbf{\overline{2}},\mathbf{2})$
& $\Lambda^{\pm}_1$  & (-3, $\mp$1, -3) \\
($R$, $\bar{R}$) & ($h_{15}$, $\phi^{-}_{15}$) & $(\mathbf{15}_2,\mathbf{\overline{2}},\mathbf{2})$
& $\Lambda^{\pm}_1$  & (-3, $\mp$1, -3) \\
\bottomrule
\end{tabular}
\end{center}
\caption{
Resulting predictions for the PS GUT scale Yukawa coupling ratios
$((Y_e)_{ij}/(Y_d)_{ij}$, $(Y_u)_{ij}/(Y_d)_{ij}$, $(Y_\nu)_{ij}/(Y_u)_{ij} )$
from the diagram in Fig.~\ref{Fig:Other}; for more details see main
text.
If the messenger representation $X$ has an index, there is more than one
way to combine the fields $A$ and $C$ or $B_1$ and $B_2$ to form this
representation.
\label{Tab:PSMessenger3}
}
\end{table}

As discussed already in the main text we have put our main
focus in this publication on diagrams like in Fig.~\ref{Fig:Dim5}
where $A$ and $C$ are GUT matter representations. For these
diagrams the messengers are matter-like (they would carry a
U(1)$_R$ charge if we would introduce an $R$-symmetry).
In fact, only this class of diagrams generates new relations
beyond the renormalizable ones.

Nevertheless, for completeness we have collected here in the
appendix also the other cases. They are described by the diagram
in Fig.~\ref{Fig:Other}. In this case the messengers are Higgs-like
($X$ would carry no $R$-charge) and their representation
determines the Yukawa coupling ratio. But since the product of
the matter fields $A$ and $C$ allows only two possible representations
containing a Higgs doublet we end up again with the two renormalizable
Yukawa coupling ratios.

Note again that we do not consider cases here where components of
the messenger pairs $X$ and $\bar X$ remain massless.


\begin{thebibliography}{10}


\bibitem{Georgi:1974sy}
  H.~Georgi and S.~L.~Glashow,
  Phys.\ Rev.\ Lett.\  {\bf 32} (1974) 438.

\bibitem{Pati:1974yy}
  J.~C.~Pati and A.~Salam,
  Phys.\ Rev.\ D {\bf 10} (1974) 275
   [Erratum-ibid.\ D {\bf 11} (1975) 703].

\bibitem{GJ}
  H.~Georgi and C.~Jarlskog,
  Phys.\ Lett.\  B {\bf 86} (1979) 297.
  
\bibitem{Ross:2007az}
  G.~Ross and M.~Serna,
  Phys.\ Lett.\ B {\bf 664} (2008) 97
  [arXiv:0704.1248 [hep-ph]].

\bibitem{Altmannshofer:2008vr}
W.~Altmannshofer, D.~Guadagnoli, S.~Raby, D.~M.~Straub,
Phys.\ Lett.\  {\bf B668 } (2008)  385-391.
[arXiv:0801.4363 [hep-ph]].

\bibitem{Baer:2009ff}
H.~Baer, S.~Kraml, A.~Lessa, S.~Sekmen,
JHEP {\bf 1002 } (2010)  055.
[arXiv:0911.4739 [hep-ph]].

\bibitem{Gogoladze:2009ug}
 I.~Gogoladze, R.~Khalid, Q.~Shafi,
 Phys.\ Rev.\  {\bf D79 } (2009)  115004.
 [arXiv:0903.5204 [hep-ph]];
 I.~Gogoladze, S.~Raza, Q.~Shafi,
 [arXiv:1104.3566 [hep-ph]];
 S.~Dar, I.~Gogoladze, Q.~Shafi, C.~S.~Un,
 [arXiv:1105.5122 [hep-ph]];
  I.~Gogoladze, Q.~Shafi and C.~S.~Un,
  JHEP {\bf 1208} (2012) 028
  [arXiv:1112.2206 [hep-ph]].
  H.~Baer, I.~Gogoladze, A.~Mustafayev, S.~Raza and Q.~Shafi,
  JHEP {\bf 1203} (2012) 047
  [arXiv:1201.4412 [hep-ph]];
  I.~Gogoladze, Q.~Shafi and C.~S.~Un,
  JHEP {\bf 1207} (2012) 055
  [arXiv:1203.6082 [hep-ph]].
  M.~Adeel Ajaib, I.~Gogoladze, Q.~Shafi and C.~S.~Un,
  JHEP {\bf 1307} (2013) 139
  [arXiv:1303.6964 [hep-ph]];
  M.~A.~Ajaib, I.~Gogoladze, Q.~Shafi and C.~S.~Un,
  arXiv:1308.4652 [hep-ph].

\bibitem{Anandakrishnan:2012tj}
  A.~Anandakrishnan, S.~Raby and A.~Wingerter,
  Phys.\ Rev.\ D {\bf 87} (2013) 5,  055005
  [arXiv:1212.0542 [hep-ph]];
  A.~Anandakrishnan, B.~C.~Bryant, S.~Raby and A.~Wingerter,
  arXiv:1307.7723 [hep-ph].

\bibitem{Antusch:2009gu}
  S.~Antusch and M.~Spinrath,
  Phys.\ Rev.\ D {\bf 79} (2009) 095004
  [arXiv:0902.4644 [hep-ph]].

\bibitem{Allanach:1996hz}
  B.~C.~Allanach, S.~F.~King, G.~K.~Leontaris and S.~Lola,
  Phys.\ Rev.\  D {\bf 56} (1997) 2632
  [arXiv:hep-ph/9610517].
  
\bibitem{Allanach:1997gu}
  B.~C.~Allanach, S.~F.~King, G.~K.~Leontaris and S.~Lola,
  Phys.\ Lett.\ B {\bf 407} (1997) 275
  [hep-ph/9703361].

\bibitem{Anderson:1993fe}
  G.~Anderson, S.~Raby, S.~Dimopoulos, L.~J.~Hall and G.~D.~Starkman,
  Phys.\ Rev.\ D {\bf 49} (1994) 3660
  [hep-ph/9308333].
  
\bibitem{Bazzocchi:2008sp}
  F.~Bazzocchi, M.~Frigerio and S.~Morisi,
  Phys.\ Rev.\ D {\bf 78} (2008) 116018
  [arXiv:0809.3573 [hep-ph]].

\bibitem{Monaco:2011wv}
  M.~Monaco and M.~Spinrath,
  Phys.\ Rev.\ D {\bf 84} (2011) 055009
  [arXiv:1106.6208 [hep-ph]].

\bibitem{Antusch:2011sq}
  S.~Antusch, L.~Calibbi, V.~Maurer and M.~Spinrath,
  Nucl.\ Phys.\ B {\bf 852} (2011) 108
  [arXiv:1104.3040 [hep-ph]].

\bibitem{Antusch:2011xz}
  S.~Antusch, L.~Calibbi, V.~Maurer, M.~Monaco and M.~Spinrath,
  Phys.\ Rev.\ D {\bf 85} (2012) 035025
  [arXiv:1111.6547 [hep-ph]].

\martin{  
\bibitem{Aulakh:2008sn}
  C.~S.~Aulakh and S.~K.~Garg,
  Nucl.\ Phys.\ B {\bf 857} (2012) 101
  [arXiv:0807.0917 [hep-ph]].
  }  
  
  
\bibitem{Antusch:2011qg}
  S.~Antusch and V.~Maurer,
  Phys.\ Rev.\ D {\bf 84} (2011) 117301
  [arXiv:1107.3728 [hep-ph]].

\bibitem{Marzocca:2011dh}
  D.~Marzocca, S.~T.~Petcov, A.~Romanino and M.~Spinrath,
  JHEP {\bf 1111} (2011) 009
  [arXiv:1108.0614 [hep-ph]].

\bibitem{Antusch:2013jca}
  S.~Antusch and V.~Maurer,
 JHEP (to appear) [arXiv:1306.6879 [hep-ph]]. 
  
\bibitem{King:2012vj}
  S.~F.~King,
  Phys.\ Lett.\ B {\bf 718} (2012) 136
  [arXiv:1205.0506 [hep-ph]].

\bibitem{Antusch:2012fb}
  S.~Antusch, C.~Gross, V.~Maurer and C.~Sluka,
  Nucl.\ Phys.\ B {\bf 866} (2013) 255
  [arXiv:1205.1051 [hep-ph]].
  
\bibitem{Antusch:2013kna}
  S.~Antusch, C.~Gross, V.~Maurer and C.~Sluka,
  arXiv:1305.6612 [hep-ph].
  
\bibitem{Antusch:2013tta}
  S.~Antusch, C.~Gross, V.~Maurer and C.~Sluka,
  arXiv:1306.3984 [hep-ph].
  
\bibitem{King:2013iva}
  S.~F.~King,
  JHEP {\bf 1307} (2013) 137
  [arXiv:1304.6264 [hep-ph]].
  
\bibitem{King:2013xba}
  S.~F.~King,
    Phys.\ Lett.\  B {\bf 724} (2013) 92
 [arXiv:1305.4846 [hep-ph]].


  

\end{thebibliography}
\end{document}